\documentclass[12pt,thmsa]{article}
\usepackage{amsmath}
\usepackage{amssymb}
\usepackage{sw20lart}

\input{tcilatex}

\begin{document}

\author{{\large F. Iddir}$\thanks{
E-mail: \emph{iddir@univ-oran.dz}}${\large \ }and {\large \ L. Semlala}$%
\thanks{
E-mail: \emph{l\_semlala@yahoo.fr}}$ \\
$Laboratoire${\small \ }$de${\small \ }$Physique${\small \ }$Th\acute{e}%
orique,${\small \ }\\
$Universit\acute{e}${\small \ }$d^{\prime }Oran${\small \ }$Es${\small -}$S%
\acute{e}nia,${\small \ }$31100,ALGERIE$}
\title{{\LARGE Heavy Hybrid Mesons Masses }}
\date{ }
\maketitle

\begin{abstract}
We estimate the ground state masses of the heavy hybrid mesons using a
phenomenological QCD-type potential. $0^{-\text{ }-},1^{-\text{ }-},0^{-%
\text{ }+},1^{-\text{ }+}$ and $0^{+\text{ }-}$ $J^{PC}$ states are
considered.
\end{abstract}

\section{Introduction}

Quantum Chromodynamics, acknowledged as the theory of strong interactions,
allows that mesons containing gluons ( \emph{q\={q}g} hybrids\ ) may exist.
The physical existence of these \textquotedblleft \emph{exotic}%
\textquotedblright\ particles (beyond the quark model) is one of the
objectives of experimental projects $^{[1]}$. These programs would
contribute significantly to the future investigation of QCD exotics, and
should improve our understanding of hybrid physics and on the role of the
gluon in QCD. From experimental efforts at IHEP $^{[2]}$, KEK $^{[3]}$, CERN 
$^{[4]}$ and BNL $^{[5]}$, several hybrid candidates have been identified,
essentially with exotic quantum numbers $J^{PC}=1^{-\text{ }+}$ at $1400$
and $1600$ $MeV.$ In the charm sector we can consider the recently observed $%
Y(4260)^{[6]}$ candidate.

We propose estimates for the masses of the hybrid mesons considering the $c$
and $b$ flavors, within the context of a quark-gluon constituent model using
a phenomenological potential motivated by QCD and taking into account some
relativistic effects.

\section{The model}

We introduce a model in which the gluon is considered as massive constituent
particle. The constituent gluon mass ($m_{g}\simeq 800$ $MeV$) is choosen as
an order of magnitude, taking in account the mass of the gluball candidate ($%
1.6$ $GeV$). Furthermore, the authors of [$7$] are generating constituent
quarks and gluon masses in the context of Dynamical Quark Model employing
BCS vacuum, and obtained a constituent gluon mass of $800$ $MeV$. As proved
in $\left[ 9\right] $, the impact of the $m_{g}$ in the results of hybrid
masses is weak and the order of magnitude remains the same.

The Hamiltonian is constructed, containing a phenomenological potential
which reproduce the QCD characteristics; its expression has the mathematical
``Coulomb + Linear'' form.

The basic hypothesis is to use a Schr\"{o}dinger-type wave equation$^{[8]}$: 
\begin{equation}
\left\{ \sum\limits_{i=1}^{N}\left( \frac{\vec{p}_{i}^{\text{ }2}}{2M_{i}}+%
\frac{M_{i}}{2}+\frac{m_{i}^{2}}{2M_{i}}\right) +V_{eff}\right\} \text{ }%
\Psi (\vec{r}_{i})\text{ }=E\text{ }\Psi (\vec{r}_{i})\text{ };  \tag{1}
\end{equation}

where $M_{i}$ are some ``dynamical masses'' satisfying the conditions: 
\begin{equation}
\frac{\partial E}{\partial M_{i}}=0\text{ };  \tag{2}
\end{equation}

$V_{eff}$ is the average over the color space of chromo-spatial potential$%
^{[15]}$: 
\begin{eqnarray}
V_{eff} &=&\left\langle V\right\rangle _{color}=\left\langle
-\sum\limits_{i<j=1}^{N}\mathbf{F}_{i}\cdot \mathbf{F}_{j}\text{ }%
v(r_{ij})\right\rangle _{color}  \TCItag{3} \\
&=&\sum\limits_{i<j=1}^{N}\alpha _{ij}v(r_{ij}).  \notag
\end{eqnarray}

The spatial term $v(r_{ij})$ is a potential motivated by QCD, which have the
form: 
\begin{equation}
v(r_{ij})=-\frac{\alpha _{s}}{r_{ij}}+\sigma \text{ }r_{ij}+c;  \tag{4}
\end{equation}

the $\alpha _{s}$, $\sigma $, and $c$ may be fitted by experimental data.

\section{The hybrid mesons}

For the classification of hybrid mesons in a constituent model we will use
the notations of $[9]:$

\emph{l}$_{\text{g}}$\emph{\ \ \ }: is the relative orbital momentum of the
gluon in the \emph{q\={q}} center of mass;

\emph{l}$_{\text{\textit{q\={q}}}}$\emph{\ \ }: is the relative orbital
momentum between \emph{q} and \emph{\={q}};

\emph{S}$_{\text{\textit{q\={q}}}}$\emph{\ }: is the total quark spin;

\emph{j}$_{\text{\textit{g}}}$\emph{\ \ \ }: is the total gluon angular
momentum;

\emph{L \ \ \ }: \textit{l}$_{\text{\textit{q\={q}}}}$ + \textit{j}$_{\text{%
\textit{g}}}.$

The parity and charge conjugation of the hybrid are given by: 
\begin{equation*}
\QDATOP{P=\left( -\right) ^{l_{q\bar{q}}+l_{g}},\text{ \ \ \ \ }}{C=\left(
-\right) ^{l_{q\bar{q}}+S_{q\bar{q}}+1}.}
\end{equation*}

We have to solve the wave equation relative to the Hamiltonian: 
\begin{equation}
H=\sum\limits_{i=q,\text{ }\bar{q},\text{ }g}\left( \frac{\vec{p}_{i}^{2}}{%
2M_{i}}+\frac{M_{i}}{2}+\frac{m_{i}^{2}}{2M_{i}}\right) +V_{eff\text{ }}; 
\tag{5}
\end{equation}

with, for the hybrid meson: 
\begin{equation*}
\QDATOP{\alpha _{q\bar{q}}=-\frac{1}{6};\text{\ \ \ \ \ \ \ \ }}{\alpha _{%
\bar{q}g}=\alpha _{qg}=\frac{3}{2}.}
\end{equation*}

We define the Jacobi coordinates: 
\begin{equation*}
\QDATOP{\vec{\rho}=\vec{r}_{\bar{q}}-\vec{r}_{q}\text{ \ \ \ \ \ \ \ \ \ \ \
\ }}{\vec{\lambda}=\vec{r}_{g}-\frac{M_{q}\vec{r}_{q}+M_{\bar{q}}\vec{r}_{%
\bar{q}}}{M_{q}+M_{\bar{q}}}}\text{ }.
\end{equation*}

Then, the relative Hamiltonian is given by: 
\begin{equation}
H_{R}=\frac{\vec{p}_{\rho }^{2}}{2\mu _{\rho }}+\frac{\vec{p}_{\lambda }^{2}%
}{2\mu _{\lambda }}+V_{eff}(\vec{\rho},\vec{\lambda})+\frac{M_{q}}{2}+\frac{%
m_{q}^{2}}{2M_{q}}+\frac{M_{\bar{q}}}{2}+\frac{m_{\bar{q}}^{2}}{2M_{\bar{q}}}%
+\frac{M_{g}}{2}+\frac{m_{g}^{2}}{2M_{g}};  \tag{6}
\end{equation}

with 
\begin{equation*}
\QDATOP{\mu _{\rho }=\left( \frac{1}{M_{q}}+\frac{1}{\text{ }M_{\bar{q}}}%
\right) ^{-1}\text{ \ \ \ \ \ }}{\mu _{\lambda }=\left( \frac{1}{M_{g}}+%
\frac{1}{M_{q}+M_{\bar{q}}}\right) ^{-1},}
\end{equation*}

and

\begin{eqnarray}
V_{eff}(\vec{\rho},\vec{\lambda}) &=&-\alpha _{s}\left( -\frac{1}{6\rho }+%
\frac{3}{2}\frac{1}{\left\vert \vec{\lambda}+\frac{\vec{\rho}}{2}\right\vert 
}+\frac{3}{2}\frac{1}{\left\vert \vec{\lambda}-\frac{\vec{\rho}}{2}%
\right\vert }\right)  \TCItag{7} \\
&&+\sigma \left( -\frac{1}{6}\rho +\frac{3}{2}\left\vert \vec{\lambda}+\frac{%
\vec{\rho}}{2}\right\vert +\frac{3}{2}\left\vert \vec{\lambda}-\frac{\vec{%
\rho}}{2}\right\vert \right) +\frac{17}{6}c+V_{S}\text{ }.  \notag
\end{eqnarray}

The followed expansion representing the hybrid wave function in the cluster
approximation: 
\begin{eqnarray}
\Psi _{J^{PC}}(\vec{\rho},\vec{\lambda}) &=&\sum\limits_{\substack{ l_{q\bar{%
q}}l_{g}j_{g}\text{ }L\text{ }  \\ (Mm\mu ) }}\varphi ^{l_{q\bar{q}}l_{g}}(%
\vec{\rho},\vec{\lambda})\text{\textbf{e}}^{\mu _{g}}\chi _{_{S_{_{q\bar{q}%
}}}}^{^{\mu _{q\bar{q}}}}\left\langle l_{g}m_{g}1\mu _{g}\mid
J_{g}M_{g}^{\prime }\right\rangle  \TCItag{8} \\
&&\times \left\langle l_{q\bar{q}}m_{q\bar{q}}J_{g}M_{g}^{\prime }\mid
Lm\right\rangle \text{ }\left\langle LmS_{q\bar{q}}\mu _{q\bar{q}}\mid J%
\text{ }0\right\rangle  \notag
\end{eqnarray}

where the summation runs over all possible values of $l_{q\bar{q}}l_{g}j_{g}$
$L$ $(Mm\mu )$ for a given $J^{PC}$ state, restricting ourselves to the
first orbital excitations ( $l_{q\bar{q}\text{, }}l_{g}\leq 2$) see Table 2,

and $\varphi ^{l_{q\bar{q}}l_{g}}(\vec{\rho},\vec{\lambda})$ are the
Gaussian wavefunctions: 
\begin{equation}
\varphi ^{l_{q\bar{q}}l_{g}}(\vec{\rho},\vec{\lambda})=\rho ^{l_{q\bar{q}%
}}\lambda ^{l_{g}}\exp \left( -\frac{1}{2}n\text{ }\beta ^{2}\text{ }\left(
\rho ^{2}+\lambda ^{2}\right) \right) \mathbf{Y}_{l_{q\bar{q}}m_{q\bar{q}%
}}(\Omega _{\rho })\mathbf{Y}_{l_{g}m_{g}}(\Omega _{\lambda });  \tag{9}
\end{equation}

The energy

\begin{equation}
E_{J^{PC}}\left( \beta ,M_{q},M_{\bar{q}},M_{g}\right) =\frac{\left\langle
J^{PC}\right| H_{R}\left| J^{PC}\right\rangle }{\langle J^{PC}\mid
J^{PC}\rangle }  \tag{10}
\end{equation}

is minimized with respect to parameters $\beta ,$ $M_{q},$ $M_{\bar{q}}$ and 
$M_{g}.$

In Table 1 we give the parameters fitting to $J^{PC}=1^{-\text{ }-}$ \textit{%
(c\={c})}\ and \textit{(b\={b})}\ spectrum.

\begin{center}
\begin{equation}
\begin{tabular}{|c|c|c|c|c|}
\hline
$\alpha _{s}$ & $\sigma $ $(GeV^{\text{ }2})$ & $c$ $(GeV)$ & $m_{c}$ $(GeV)$
& $m_{b}$ $(GeV)$ \\ \hline
$0.36$ & $0.144$ & $-0.45$ & $1.70$ & $5.05\text{ }$ \\ \hline
\end{tabular}
\tag{\QTR{it}{Table\ 1:\ Heavy flavors potential parameters.}}
\end{equation}
\end{center}

We present in Table 2 our estimates of hybrid mesons masses for different $%
J^{PC}$ states, we take $800$ $MeV$ \ for the mass of the gluon.

\begin{center}
\begin{equation}
\begin{tabular}{|c|c|c|c|c|c|c|c|}
\hline
$J^{PC}$ & $S_{q\bar{q}}$ & $l_{q\bar{q}}$ & $l_{g}$ & $j_{g}$ & $L$ & $M_{c%
\overline{c}g}$ & $M_{b\overline{b}g}$ \\ \hline
$0^{-\text{ }-}$ & $%
\begin{array}{l}
0 \\ 
1%
\end{array}%
$ & $%
\begin{array}{l}
0,2 \\ 
1%
\end{array}%
$ & $%
\begin{array}{l}
1 \\ 
0,2%
\end{array}%
$ & $%
\begin{array}{l}
0,2 \\ 
1,2%
\end{array}%
$ & $%
\begin{array}{l}
0 \\ 
1%
\end{array}%
$ & $4.73$ & $11.02$ \\ \hline
$1^{-\text{ }-}$ & $%
\begin{array}{l}
0 \\ 
1%
\end{array}%
$ & $%
\begin{array}{l}
0,2 \\ 
1%
\end{array}%
$ & $%
\begin{array}{l}
\text{ }1 \\ 
0,2%
\end{array}%
$ & $%
\begin{array}{l}
\text{ }1 \\ 
1,2,3%
\end{array}%
$ & $%
\begin{array}{l}
1 \\ 
0,1,2%
\end{array}%
$ & $4.82$ & $11.12$ \\ \hline
$0^{-\text{ }+}$ & $%
\begin{array}{l}
0 \\ 
1%
\end{array}%
$ & $%
\begin{array}{l}
1 \\ 
0,2%
\end{array}%
$ & $%
\begin{array}{l}
0,2 \\ 
1%
\end{array}%
$ & $%
\begin{array}{l}
1 \\ 
1,2%
\end{array}%
$ & $%
\begin{array}{l}
0 \\ 
1%
\end{array}%
$ & $4.70$ & $10.96$ \\ \hline
$1^{-\text{ }+}$ & $%
\begin{array}{l}
0 \\ 
1%
\end{array}%
$ & $%
\begin{array}{l}
1 \\ 
0,2%
\end{array}%
$ & $%
\begin{array}{l}
0,2 \\ 
1%
\end{array}%
$ & $%
\begin{array}{l}
1,2 \\ 
0,1,2%
\end{array}%
$ & $%
\begin{array}{l}
1 \\ 
0,1,2%
\end{array}%
$ & $4.72$ & $10.98$ \\ \hline
$0^{+\text{ }-}$ & $%
\begin{array}{l}
0 \\ 
1%
\end{array}%
$ & $%
\begin{array}{l}
2 \\ 
1%
\end{array}%
$ & $%
\begin{array}{l}
2 \\ 
1%
\end{array}%
$ & $%
\begin{array}{l}
2 \\ 
0,1,2%
\end{array}%
$ & $%
\begin{array}{l}
0 \\ 
1%
\end{array}%
$ & $4.58$ & $10.86$ \\ \hline
\end{tabular}
\tag{\QTR{it}{Table 2: Heavy hybrid mesons quantum numbers and
masses
estimations (in GeV)}}
\end{equation}
\end{center}

We compare our results with recent LQCD predictions$^{[10]}$(Table 3).

\begin{equation}
\begin{tabular}{|l|l|}
\hline
$J^{PC}$ & $M_{c\overline{c}g}$ \\ \hline
$1^{-\text{ }+}$ & $4.40$ \\ \hline
$0^{+\text{ }-}$ & $4.73$ \\ \hline
$0^{-\text{ }-}$ & $5.88$ \\ \hline
\end{tabular}
\tag{\QTR{it}{Table 3:Recent LQCD hybrid predictions}$^{[10]}$\QTR{it}{.}}
\end{equation}

For the orders of magnitude of the masses, our results are in agreement with
the recent LQCD predictions$^{[10]}$(Table 3), except the case of $0^{-\text{
}-}$state which is very more heavier than our results.

\section{References}

[1] \ \ T. Barnes, nucl-th/9907020; M. A. Moinester and V. Steiner,
hep-ex/9801011; N. N. Achasov and G. N. Shestakov, hep-ph/9901380; G. R.
Blackett et \textit{al, }hep-ex/9708032; D. Lie, H. Yu and Q. X. Shen,
hep-ph/0001063; A. V. Afanasev and A. P. Szczepaniac, hep-ph/9910268

[2] \ \ D. Alde et \textit{al, }Proc. of Hadron 97; Yu D. Prokoshin and S.
A. Sadorsky, Phys. At. Nucl. \textbf{58} (1995) 606; G. M. Beladidze et 
\textit{al, }Phys. Lett. \textbf{B313} (1993) 276; A. Zaitsev, Proc. of
Hadron 97

[3] \ \ H. Aoyagi et \textit{al, }Phys. Lett. \textbf{B314} (1993) 246

[4] \ \ D. Alde et \textit{al,} Phys. Lett. \textbf{B205} (1988) 397

[5] \ \ D. R. Thompson et \textit{al,} (E852Coll.), Phys. Rev. Lett.\textbf{%
\ 79} (1997) 1630

[6] Tadeusz Lesiak, hep-ex/0511003; B. Aa. Petersen, hep-ex/0609030.

[7] \ \ E. S. Swanson and A. P. Szczepaniak, hep-ph/9804219; A. P.
Szczepaniak, E. S. Swanson, C.-R. Ji and S.R. Cotanch, Phys. Rev. Lett.%
\textbf{\ 76}, 2011(1996).

[8] \ H. G. Dosch, Yu. A. Simonov, Phys. Lett. \textbf{B205}, 339 (1988);
Yu. A. Simonov, Phys. Lett. \textbf{B226}, 151 (1989), ibid. \textbf{B228},
413 (1989); Yu. A. Simonov, Heidelberg Preprint HD THEP-91-5 (1991)

[9] \ \ F. Iddir and L. Semlala, hep-ph/0211289

[10] \ \ Yan Liyu and Xiang-Qian Luo, hep-Lat/0511015

\end{document}